\documentclass[final]{raa_twocolumn}
\usepackage{graphicx,times}
\usepackage{natbib}
\usepackage{algpseudocode}
\usepackage{algorithm}
\usepackage{amssymb,amsmath}
\usepackage{lipsum,amsmath,multicol}

\bibpunct{(}{)}{;}{a}{}{,}

\usepackage{hyperref}
\hypersetup{pdftitle = The title of my PDF, pdfauthor = My name, pdfsubject= The subject, pdfkeywords = keyword1 keyword2 keyword3}
\hypersetup{colorlinks = true, linkcolor = green, anchorcolor = red, citecolor = blue, filecolor = red, pagecolor = red, urlcolor = red}

\begin{document}
\title{Symplectic Integrators in Corotating Coordinates}

\setcounter{page}{1}

\author{Xiongbiao Tu$^1$, Qiao Wang$^{1,4}$, Yifa Tang$^{2,3}$}

\institute{
$^1$National Astronomical Observatories, Chinese Academy of Sciences, Beijing 100101, China\\
$^2$LSEC, ICMSEC, Academy of Mathematics and Systems Science, Chinese Academy of Sciences, Beijing 100190, China\\
$^3$School of Mathematical Sciences, University of Chinese Academy of Sciences, Beijing 100049, China\\
$^4$School of Astronomy and Space Science, University of Chinese Academy of Sciences, Beijing 100049, China \\
$qwang@nao.cas.cn$
\vs \no
{\small Received 0000 xxxx 00; accepted 0000 xxxx 00}
}

\abstract{
The dynamic equation of mass point in rotating coordinates is governed by Coriolis and centrifugal force, besides a corotating potential relative to frame. Such a system is no longer a canonical Hamiltonian system so that the construction of symplectic integrator is problematic. In this paper, we present three integrators for this question. It is significant that those schemes have the good property of near-conservation of energy.  We proved that the discrete  symplectic map of  $(\mathbf{p}_{n},\mathbf{x}_{n}) \mapsto (\mathbf{p}_{n+1},\mathbf{x}_{n+1})$ in corotating coordinates exists and the two integrators are variational symplectic. Two groups of numerical experiments demonstrates the precision and long-term convergence of these integrators in the examples of corotating top-hat density and  circular restricted three-body system.
}

\authorrunning{X. Tu, Q. Wang \& Y. Tang}  
\titlerunning{Symplectic Integrators in Corotating Coordinates}
\maketitle
\section{Introduction}

 
Canonical Hamiltonian system could be the most important physical systems 
and a canonical Hamiltonian system in the variables $\mathbf{z} = (p,q)$ given in the form
\begin{equation}\label{eq0.1}
\begin{array}{l}
\dot{p}=- H_{q}(p,q) \\[1mm]
\dot{q}=~~~H_{p}(p,q)
\end{array}
\end{equation}
where $p,q \in \mathbb{R}^{d}$. Or equivalently
\begin{equation}\label{eq0.2}
\dot{\mathbf{z}}=J^{-1}\nabla H(\mathbf{z}),
~
J= \left(\begin{array}{ll}
~~~0 &~ I_{d}\\
-I_{d}  &~ 0
\end{array}\right),
\end{equation}
where $\mathbf{z} = (\mathbf{x},\mathbf{v})^{\top}$ and $I_{d}$ is a $d \times d$ identity matrix. It has an outstanding property that the flow of Hamiltonian system is symplectic. It is natural to find those discrete systems which preserve the properties of symplecticity and the inner symmetries of original Hamiltonian system. So the symmetric, symplectic algorithms \citep{Feng1985,Feng1986,Forest1990,Channell1990,Candy1991} are the standard methods for such problems. 

 For example, the implicit midpoint scheme, which is symmetric, second-order, symplectic integrator for the canonical Hamiltonian system. The other example is the well known Boris algorithm \citep{Boris1970} in the plasma dynamics, which has some good geometric properties. Generally, it is symmetric, second-order, volume-preserving \citep{Qin2013PhPl}, and is not symplectic \citep{ELLison2015JCoP}. But, in special configuration of the homogeneous magnetic field, the integrator is variational symplectic and preserve near-conservation of energy over long term evolution\citep{ELLison2015JCoP,Hairer2018BIT}. 

Unfortunately, the motion of mass points in corotating frame is a non-canonical Hamiltonian system. So it is not available to construct symplectic numerical methods in the direct approach. However, the calculation of precise numerical orbit is critical for the dynamic studies of binary star system, central bar in galaxies, etc \citep{2008gady.book.....B}. Even considering a simple restricted three body problem, such as halo orbits about Lagrange points, it is non-trivial to find a high precise orbit in Earth-Moon corotating coordinate \citep{Akiyama2019AcAau, Oshima2019CeMDA}. In this work, we construct and investigate three integrators for the geometric properties or conservation in corotating potentials. 
This paper is organized as follows. In Sec. 2, we give a brief introduction to corotating coordinate system and three numerical methods. In Sec. 3 and 4, we analysis the long time energy behaviours for these numerical methods and show that these numerical methods have some good geometric properties. In Sec. 5, two groups of numerical experiments were performed to check the precision and demonstrate good geometric properties. Finally, we summarize this work in Sec. 6.

\section{Numerical methods}
\subsection{The corotating system}
The equations of motion in corotating coordinates can be written as 
\begin{equation}\label{rotate}
\begin{array}{l}
\ddot{\mathbf{x}} + 2(\Omega\times\dot{\mathbf{x}})= -\nabla(U(\mathbf{x})-\frac{1}{2}\omega^2\mathbf{r}^2),
\end{array}
\end{equation}
 where $\mathbf{x}=(x,y,z)$ is the position, $\Omega=(0,0,\omega)$ means the system rotates clockwise around the $z$ axis with rotation speed $\omega$, $U(\mathbf{x})$ is a potential energy, and $\mathbf{r}=(x,y,0)$. It is an Euler-Lagrange equations with Lagrangian $L(\mathbf{x},\dot{\mathbf{x}})=\frac{1}{2}\dot{\mathbf{x}}^2 + \Omega\cdot (\mathbf{r}\times\dot{\mathbf{r}}) - (U(\mathbf{x})-\frac{1}{2}\omega^2\mathbf{r}^2)$, the conjugate momenta $\mathbf{p} = {\partial L}/{\partial \dot{\mathbf{x}}} = \dot{\mathbf{x}} + (-\omega y,\omega x,0)$ (conjugate to the position variables $\mathbf{x}$) derived by Legendre transform. The energy $E = \frac{1}{2}\dot{\mathbf{x}}^2 + (U(\mathbf{x})-\frac{1}{2}\omega^2\mathbf{r}^2)$ is an invariant along the flow of the system.

We set $\varphi(\mathbf{x}) = U(\mathbf{x})-\frac{1}{2}\omega^2\mathbf{r}^2$ and rewrite the corotating coordinate system (\ref{rotate}). Let $\dot{\mathbf{x}} = \mathbf{v}$, $\mathbf{z} = (\mathbf{x},\mathbf{v})^{\top}$, and $\mathbf{v}$ is the velocity of the particle. The motion equations of the particle can be expressed as 
\begin{equation} \label{rerotating}
\left\{\begin{aligned}
\dot{\mathbf{x}} &= \mathbf{v},\\
\dot{\mathbf{v}} &= 2\mathbf{v}\times\Omega - \nabla\varphi(\mathbf{x}).
\end{aligned}\right.
\end{equation}
Obviously, it is a non-canonical Hamiltonian system $\dot{\mathbf{z}}= K^{-1}\nabla H(\mathbf{z})$ with
\begin{equation}\notag
K=\left(\begin{array}{cc}
    \hat{\Omega} &  -I_3\\
    I_3 & 0
\end{array}\right), \quad
\hat{\Omega}=\left(\begin{array}{ccc}
    0 & 2\omega & 0\\
    -2\omega & 0 & 0\\
    0 & 0 & 0
\end{array}\right).
\end{equation}
Here, $K$ is an antisymmetric matrix with its entries being the rotation speed of the system rotated, and $H(\mathbf{z})= \frac{1}{2}\mathbf{v}^2 + \varphi(\mathbf{x})$.
The antisymmetric matrix $K$ provides a K-symplectic structure, which is defined by 
\begin{equation}\notag
w_K=\frac{1}{2}d\mathbf{z}^{\top}\wedge Kd\mathbf{z}.    
\end{equation}
An integrator $\psi:\mathbf{z}_n\mapsto\mathbf{z}_{n+1}$ is referred to as K-symplectic, when $w_K$ is preserved by the flow of the integrator, i.e., $d\mathbf{z}_{n+1}^{\top}\wedge Kd\mathbf{z}_{n+1}=d\mathbf{z}_n^{\top}\wedge Kd\mathbf{z}_n$.

\subsection{The Integrators}
Fisrt, we introduce the implicit midpoint scheme $\psi_m$ in coratoting coordinates. It reads
\begin{equation} \label{midpoint}
\mathbf{z}_{n+1} = \mathbf{z}_{n} + \Delta t f(\frac{\mathbf{z}_{n}+\mathbf{z}_{n+1}}{2}).
\end{equation}
$\psi_m$ is a second-order implicitly symmetric scheme. It is well-known that it is only symplectic in canonical Hamiltonian system. Generally speaking, $\psi_m$ should have been not symplectic in our non-canonical case. However we show $\psi_m$ is indeed symplectic for the system (\ref{rerotating}) in the Section 4. 

We take into account of Boris algorithm \citep{Boris1970} to discrete the system (\ref{rotate}) as the numerical integrator $\psi_b$:
\begin{equation}\label{psi1}
\begin{array}{l}
\dfrac{\mathbf{x}_{n+1}-2\mathbf{x}_{n}+\mathbf{x}_{n-1}}{\Delta t^2} +\\[0.5mm] 2\Omega\times\dfrac{\mathbf{x}_{n+1}-\mathbf{x}_{n-1}}{2\Delta t}= -\nabla\varphi(\mathbf{x}_n).
\end{array}
\end{equation}
It is a second-order explicitly symmetric integrator. At the same time, we set the discrete velocity has the form of 
\begin{equation}\label{}
\begin{split}
\mathbf{v}_{n} =& \dfrac{\mathbf{x}_{n+1} - \mathbf{x}_{n}}{\Delta t} - \frac{1}{2}(\mathbf{x}_{n+1} - \mathbf{x}_{n})\times \Omega \\
&+ \Delta t\nabla \varphi(\mathbf{x}_{n}),\\
\mathbf{v}_{n+1} =& \dfrac{\mathbf{x}_{n+1} - \mathbf{x}_{n}}{\Delta t} + \frac{1}{2}(\mathbf{x}_{n+1} - \mathbf{x}_{n})\times \Omega.
\end{split}
\end{equation}
The map $(\mathbf{v}_{n},\mathbf{x}_{n})\mapsto(\mathbf{v}_{n+1},\mathbf{x}_{n+1})$ is K-symplectic which will be verifyed in the Section 4.


Similarly, we discrete the system (\ref{rotate}) on the velocity term by the numerical integrator $\psi_v$:
\begin{equation}\label{psi2}
\begin{array}{l}
\dfrac{\mathbf{x}_{n+2} -2\mathbf{x}_{n} + \mathbf{x}_{n-2}}{(2\Delta t)^2} + 2\Omega\times\\[1mm]
\dfrac{-\mathbf{x}_{n+2}+8\mathbf{x}_{n+1}-8\mathbf{x}_{n-1}+ \mathbf{x}_{n-2}}{12\Delta t}=-\nabla\varphi(\mathbf{x}_n).\\[3mm]
\end{array}
\end{equation}
This numerical integrator $\psi_v$ is a simple modify for $\psi_b$ by using five points difference approximate for $\dot{\mathbf{x}}$. It is an explicitly symmetric, second-order numerical method.

In the next sections, we analysis the energy errors over long times and the geometry properties of these numerical methods.

\section{Energy error analysis}

In this section, we analyse the energy deviation of the integrators $\psi_b, \psi_v$ over very long times. Firstly, we consider $\psi_b$ and solve the modified differential equation whose solution $\mathbf{y}(t)$ formally satisfies $\mathbf{y}(n\Delta t)= \mathbf{x}_{n}$. Thus, $\mathbf{y}(t)$ must satisfies Eq.~(\ref{psi1}), i.e.,
\begin{equation}\label{}
\begin{array}{l}
\dfrac{\mathbf{y}(t+\Delta t)-2\mathbf{y}(t)+\mathbf{y}(t-\Delta t)}{\Delta t^2} +\\[2mm] 2\Omega\times\dfrac{\mathbf{y}(t+\Delta t)-\mathbf{y}(t-\Delta t)}{2\Delta t}= -\nabla\varphi(\mathbf{y}(t)).
\end{array}
\end{equation}
We expand all terms into powers of $\Delta t$ at the time $t$ then obtain the following modified differential equation
\begin{equation}\label{MDE}
\begin{array}{l}
(\ddot{\mathbf{y}} + \frac{\Delta t^2}{12}\mathbf{y}^{(4)} + \dots) +\\[1mm]
\Omega\times (\dot{\mathbf{y}} + \frac{\Delta t^2}{3}\mathbf{y}^{(3)}+ \dots)= -\nabla\varphi(\mathbf{y}).
\end{array}
\end{equation}
Multiplying $\dot{\mathbf{y}}^{\top}$ in the two sides of the formula. Since $\dot{\mathbf{y}}^{\top}(\Omega\times \dot{\mathbf{y}})=0$, we derive 
\begin{equation}\label{}
\begin{array}{l}
\dot{\mathbf{y}}^{\top}(\ddot{\mathbf{y}} + \frac{\Delta t^2}{12}\mathbf{y}^{(4)} + \dots) +\\[1mm] \dot{\mathbf{y}}^{\top}\Omega\times (\frac{\Delta t^2}{3}\mathbf{y}^{(3)}+ \dots)= -\dot{\mathbf{y}}^{\top}\nabla\varphi(\mathbf{y}).
\end{array}
\end{equation}
The left hand side can be written as the full differential and $\dot{\mathbf{y}}^{\top}\nabla\varphi(\mathbf{y}(t))=\frac{d}{dt}\varphi(\mathbf{y}(t))$, so the modified differential equation has a formal invariant, i.e.,
\begin{equation}\label{invar}
\begin{split}
\dfrac{d}{dt}\Big(&\frac{1}{2}\dot{\mathbf{y}}^{\top}\dot{\mathbf{y}} + \varphi(\mathbf{y}) + \frac{\Delta t^2}{12}(\dot{\mathbf{y}}^{\top}\mathbf{y}^{3}-\frac{1}{2}\ddot{\mathbf{y}}^{\top}\ddot{\mathbf{y}}\\ 
&+4\dot{\mathbf{y}}^{\top}(\Omega\times\ddot{\mathbf{y}})) +\dots\Big) = 0.
\end{split}
\end{equation}

Thus, we obtain a new formal generalized energy $E_h(\mathbf{y},\dot{\mathbf{y}}) = E(\mathbf{y},\dot{\mathbf{y}}) + \Delta t^2E_2(\mathbf{y},\dot{\mathbf{y}}) + \dots$, which is an invariant. We only consider the numerical integrator $\{(\mathbf{x}_n,\dot{\mathbf{x}}_n)\}$ in a compact set $D$. In order to estimate the energy error of the integrator $\psi_b$ (and $\psi_v$) over a long time, we truncate the $E_h$ in $N$ leading terms, and integrate over the time interval $[0, n\Delta t]$,
\begin{equation}\label{esti1}
\begin{array}{l}
E_h^N(\mathbf{x}_n,\dot{\mathbf{x}}_n) - E_h^N(\mathbf{x}_0,\dot{\mathbf{x}}_0) = n\Delta t\mathcal{O}(\Delta t^N).
\end{array}
\end{equation}
The right hand side in the formula (\ref{esti1}) is a high order infinitesimal quantity, so 
\begin{equation}\label{esti2}
\begin{array}{l}
|E(\mathbf{x}_n,\dot{\mathbf{x}}_n) - E(\mathbf{x}_0,\dot{\mathbf{x}}_0)| \leq C_N\Delta t^2.
\end{array}
\end{equation}
where $C_N$ is directly dependent on the value of $\dot{\mathbf{y}}^{\top}\mathbf{y}^{3}-\frac{1}{2}\ddot{\mathbf{y}}^{\top}\ddot{\mathbf{y}} +4\dot{\mathbf{y}}^{\top}(\Omega\times\ddot{\mathbf{y}})$ in the compact set $D$.

For the integrator $\psi_v$ in formula (\ref{psi2}), its modified differential equation as follow
\begin{equation}\label{MDEv}
\begin{array}{l}
(\ddot{\mathbf{y}} + \frac{\Delta t^2}{3}\mathbf{y}^{(4)} + \dots) +\\[1mm]
\Omega\times (\dot{\mathbf{y}} - \frac{\Delta t^4}{15}\mathbf{y}^{(5)}+ \dots)
= -\nabla\varphi(\mathbf{y}).
\end{array}
\end{equation}
By the same way, we can also obtain the energy error estimation (\ref{esti2}) of $\psi_v$.

\section{The symplectic property}
For any hyperregular Lagrangian $L(\mathbf{x},\dot{\mathbf{x}})$, the Euler-Lagrange equations are equivalent to Hamilton's equations of motion. In canonical Hamiltonian system, a map $\phi:\mathbf{z}_{n}\mapsto \mathbf{z}_{n+1}, \mathbf{z}\in \mathbb{R}^{2d}$ is called symplectic if its Jacobian matrix satisfies the symplectic condition,
\begin{equation}\label{symp}
\Big{(}\dfrac{\partial\phi}{\partial\textbf{z}_{n}}\Big{)}^{\top}J\Big{(}\dfrac{\partial\phi}{\partial\textbf{z}_{n}}\Big{)}=J.
\end{equation}
The equivalent expression is that the map $\phi$ preserves a standard symplectic structure $\frac{1}{2}d\mathbf{z}^{\top}\wedge Jd\mathbf{z}$, i.e., $d\mathbf{z}_{n+1}^{\top}\wedge Jd\mathbf{z}_{n+1}=d\mathbf{z}_n^{\top}\wedge Jd\mathbf{z}_n$.

\subsection{The symplectic property -- the implicit midpoint scheme $\psi_m$}
We discrete the system (\ref{rerotating}) by the implicit midpoint scheme (\ref{midpoint}) and rewrite in the form of only variable $\mathbf{x}$, as follow
\begin{equation}\label{disrerotating}
\begin{array}{l}
\dfrac{\mathbf{x}_{n+1}-2\mathbf{x}_{n}+\mathbf{x}_{n-1}}{\Delta t^2} + 2\Omega\times\dfrac{\mathbf{x}_{n+1}-\mathbf{x}_{n-1}}{2\Delta t} =\\[2mm] -\frac{1}{2}\!\left(\nabla\varphi\!\left(\dfrac{\mathbf{x}_{n+1}+\mathbf{x}_{n}}{2}\right) + \nabla\varphi\!\left(\dfrac{\mathbf{x}_{n}+\mathbf{x}_{n-1}}{2}\right)\right).
\end{array}
\end{equation}
Considering Lagrangian $L(\mathbf{x},\dot{\mathbf{x}})=\frac{1}{2}\dot{\mathbf{x}}^2 + \Omega\cdot (\mathbf{r}\times\dot{\mathbf{r}}) - \varphi(\mathbf{x})$ of the corotating coordinate system, we derive the discrete form of $\int_{t_n}^{t_{n+1}}L(\mathbf{x}(t),\dot{\mathbf{x}}(t))dt$. The action $S_h$ is
\begin{equation}\label{}
S_h(\mathbf{x}_0,\dots,\mathbf{x}_{N}) =\sum_{n=0}^{N-1}L_h(\mathbf{x}_n,\mathbf{x}_{n+1}),
\end{equation}
where the $L_h$ is discrete Lagrangian. One of differential form can be written as
\begin{equation}\label{eq:lag}
\begin{array}{l}
L_h(\mathbf{x}_n,\mathbf{x}_{n+1})=\frac{1}{2\Delta t}(\mathbf{x}_{n+1}-\mathbf{x}_{n})^2 + \\[1mm]
\frac{1}{2}\Omega\!\cdot\!(\mathbf{r}_n + \mathbf{r}_{n+1})\!\times\!(\mathbf{r}_{n+1}\!-\!\mathbf{r}_n)\!-\!\Delta t\varphi(\frac{\mathbf{x}_{n}\!+\!\mathbf{x}_{n+1}}{2}),
\end{array}
\end{equation}
the term $\dot{\mathbf x}$ is replaced by  $(\mathbf{x}_{n+1}-\mathbf{x}_{n})/\Delta t$. 
According to the discrete Hamilton's principle, the discrete Euler-Lagrange equation reads
\begin{equation}\label{disEuler}
\begin{array}{l}
D_2 L_h(\mathbf{x}_{n-1},\mathbf{x}_{n}) + D_1 L_h(\mathbf{x}_{n},\mathbf{x}_{n+1}) = 0,
\end{array}
\end{equation}
where $D_i$ is the partial derivative with respect to the i-th argument. The equation of motion governed by the  Lagrangian (Eq.~\ref{eq:lag}) is exactly identical with the equation of midpoint scheme (Eq.~\ref{disrerotating}). In addition, the discrete conjugate momenta is  defined by 
\begin{equation}\label{varsymp}
\mathbf{p}_{n} = -D_1 L_h(\mathbf{x}_{n},\mathbf{x}_{n+1}).
\end{equation}
Using Eq.~\ref{disEuler}, one can obtain  $\mathbf{p}_{n+1} = D_2 L_h(\mathbf{x}_{n},\mathbf{x}_{n+1})$. A straightforward  calculation gives the equation of $d\mathbf{p}_{n+1}\wedge d\mathbf{x}_{n+1}= d\mathbf{p}_{n}\wedge d\mathbf{x}_{n}$ (refer to the Theorem 5.1 of Chapter VI in \citet{Hairer2006}). That means the map $(\mathbf{p}_{n},\mathbf{x}_{n}) \mapsto (\mathbf{p}_{n+1},\mathbf{x}_{n+1})$ is symplectic and integrator $\psi_m$ is variational symplectic.

Further, substituting the relation of $\mathbf{p} = {\partial L}/{\partial \dot{\mathbf{x}}} = \mathbf{v} + (-\omega y,\omega x,0)$, one can easily compute the map of $(\mathbf{v}_{n},\mathbf{x}_{n}) \mapsto (\mathbf{v}_{n+1},\mathbf{x}_{n+1})$ (refer to the formula (6) in \citet{Tu2016PhPl}), which is a K-symplectic integrator.
So the  $\psi_m$ also implies a K-symplectic integrator
\begin{equation}\label{}
\begin{split}
\mathbf{v}_{n} = &\dfrac{\mathbf{x}_{n+1} - \mathbf{x}_{n}}{\Delta t} - (\mathbf{x}_{n+1} - \mathbf{x}_{n})\times \Omega \\[1mm] 
&+\dfrac{\Delta t}{2}\nabla \varphi(\dfrac{\mathbf{x}_{n+1}+\mathbf{x}_{n}}{2}),\\
\mathbf{v}_{n+1} = &\dfrac{\mathbf{x}_{n+1} - \mathbf{x}_{n}}{\Delta t} + (\mathbf{x}_{n+1} - \mathbf{x}_{n})\times \Omega \\[1mm] 
&-\dfrac{\Delta t}{2}\nabla \varphi(\dfrac{\mathbf{x}_{n+1}+\mathbf{x}_{n}}{2}).
\end{split}
\end{equation}


It is well known that the symplectic integrator has the property of near-conservation of energy over long times. The  error is estimated by the form (\ref{esti2}) relative to the formal energy of the modified equation \citep{Tang199431}.

\subsection{The symplectic property -- $\psi_b$}
In this subsection, we show that the numerical method $\psi_b$ is also symplectic.
By the same way, we choose the discrete Lagrangian $L_h$ as
\begin{equation}\label{}
\begin{array}{l}
L_h(\mathbf{x}_n,\mathbf{x}_{n+1})=\frac{1}{2\Delta t}(\mathbf{x}_{n+1}-\mathbf{x}_{n})^2
+\\
\frac{1}{2}\Omega\cdot (\mathbf{r}_n + \mathbf{r}_{n+1})\times(\mathbf{r}_{n+1}-\mathbf{r}_n)- \Delta t\varphi(\mathbf{x}_{n}).
\end{array}
\end{equation}
Similarly, we use the Euler-Lagrangian equation and obtain the discrete equation of
\begin{equation}\label{disEulerb}
\begin{array}{l}
D_2 L_h(\mathbf{x}_{n-1},\mathbf{x}_{n}) + D_1 L_h(\mathbf{x}_{n},\mathbf{x}_{n+1}) = 0,
\end{array}
\end{equation}
\begin{equation}\label{varsympb}
\begin{array}{l}
\mathbf{p}_{n} = -D_1 L_h(\mathbf{x}_{n},\mathbf{x}_{n+1}),\\ 
\mathbf{p}_{n+1} = D_2 L_h(\mathbf{x}_{n},\mathbf{x}_{n+1}).
\end{array}
\end{equation}
The corresponding equation of motion has the Boris form of
\begin{equation}\label{}
\begin{array}{l}
\dfrac{\mathbf{x}_{n+1}-2\mathbf{x}_{n}+\mathbf{x}_{n-1}}{\Delta t} +\\[2mm] 2\Omega\times\dfrac{\mathbf{x}_{n+1}-\mathbf{x}_{n-1}}{2} = -\Delta t\nabla\varphi(\mathbf{x}_{n}).
\end{array}
\end{equation}
It proves that the integrator $\psi_b$ is symplectic and defines a 
symplectic map $(\mathbf{p}_{n},\mathbf{x}_{n})\mapsto(\mathbf{p}_{n+1},\mathbf{x}_{n+1})$. Considering the map $(\mathbf{v}_{n},\mathbf{x}_{n}) \mapsto (\mathbf{v}_{n+1},\mathbf{x}_{n+1})$, one can substitute the relation $\mathbf{p} = {\partial L}/{\partial \dot{\mathbf{x}}} = \mathbf{v} + (-\omega y,\omega x,0)$ and derive a K-symplectic numerical integrator as 
\begin{equation}\label{}
\begin{split}
\mathbf{v}_{n} = &\dfrac{\mathbf{x}_{n+1} - \mathbf{x}_{n}}{\Delta t} - (\mathbf{x}_{n+1} - \mathbf{x}_{n})\times \Omega \\
&+ \Delta t\nabla \varphi(\mathbf{x}_{n}),\\
\mathbf{v}_{n+1} =& \dfrac{\mathbf{x}_{n+1} - \mathbf{x}_{n}}{\Delta t} + (\mathbf{x}_{n+1} - \mathbf{x}_{n})\times \Omega.
\end{split}
\end{equation}


\section{Numerical experiments}
In this section, we numerically present the behaviors of the integrators $\psi_b$, $\psi_v$ and $\psi_m$ in two kinds of extreme potential energy, a extensive Quadratic potential and restricted three-body Earth-Moon system. The reference orbits are computed by RK3 (third-order Runge-Kutta method). Specifically, the form of RK3 reads
\begin{equation}
   \left\{\begin{aligned}
   &\mathbf{z}_{n+1} = \mathbf{z}_{n} + \frac{\Delta t}{2}\left[f(K_1) + f(K_2)\right], \\
   &K_1 = \mathbf{z}_{n} + \frac{\Delta t}{6}\left[3f(K_1) - \sqrt{3}f(K_2)\right],\\
   &K_2 = \mathbf{z}_{n} + \frac{\Delta t}{6}\left[\sqrt{3}f(K_1) + 3f(K_2)\right].
   \end{aligned}\right.
\end{equation}

\subsection{Quadratic potential}
We consider a homogeneous rotating  top-hat density sphere with the quadratic potential of $U(\mathbf{x})=4(x^2+y^2+z^2)$. 
We set the rotating speed $\omega = \pi/40$ and the period is $80$. The initial position and velocity is $\mathbf{x}=(-1.9,0,0)$ and $\mathbf{v}=(0,-1.0,0)$, respectively. The time step is fixed to $\Delta t = 0.02$.

\begin{figure*}[htb]
  \centering
	\includegraphics[width=0.45\linewidth]{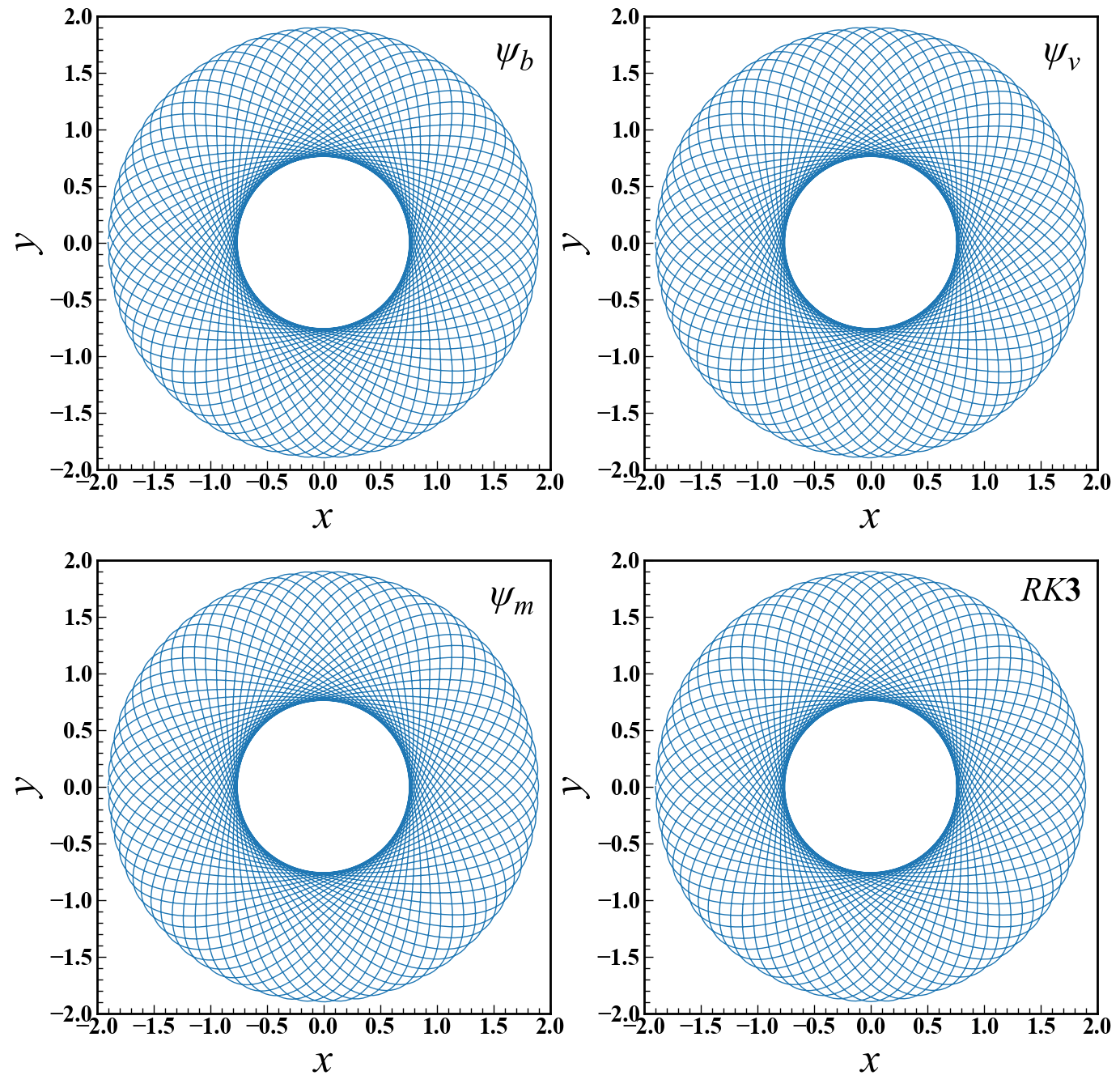}
	\includegraphics[width=0.45\linewidth]{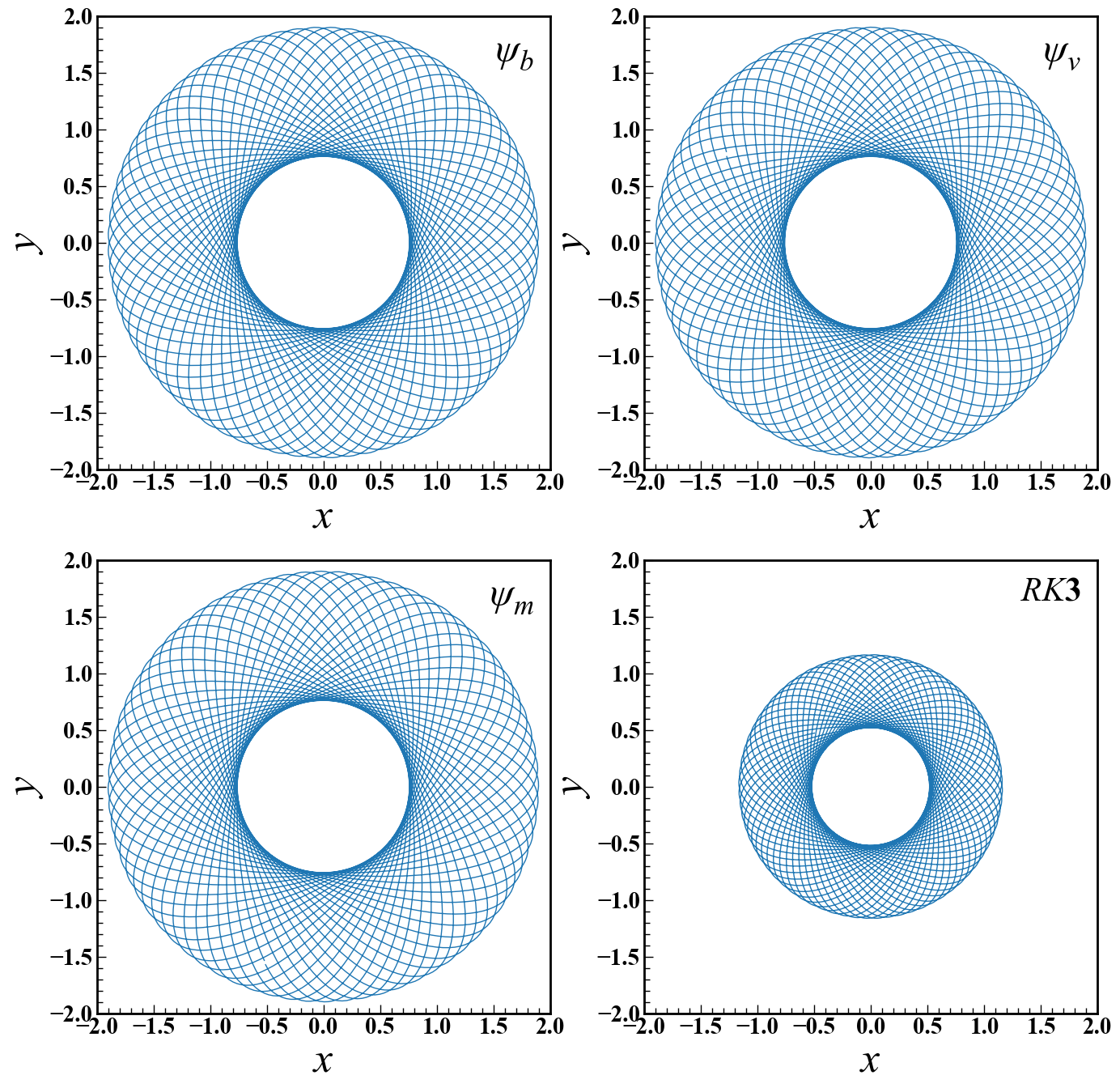}
  \caption{The orbits are given by $\psi_b, \psi_v$, the midpoint scheme $\psi_m$ and RK3 in the first rotation period (left) and the 300th rotation period (right).}
  \label{fig1.1}
\end{figure*}

\begin{figure*}[htb]
  \centering
  \includegraphics[width=0.8\linewidth]{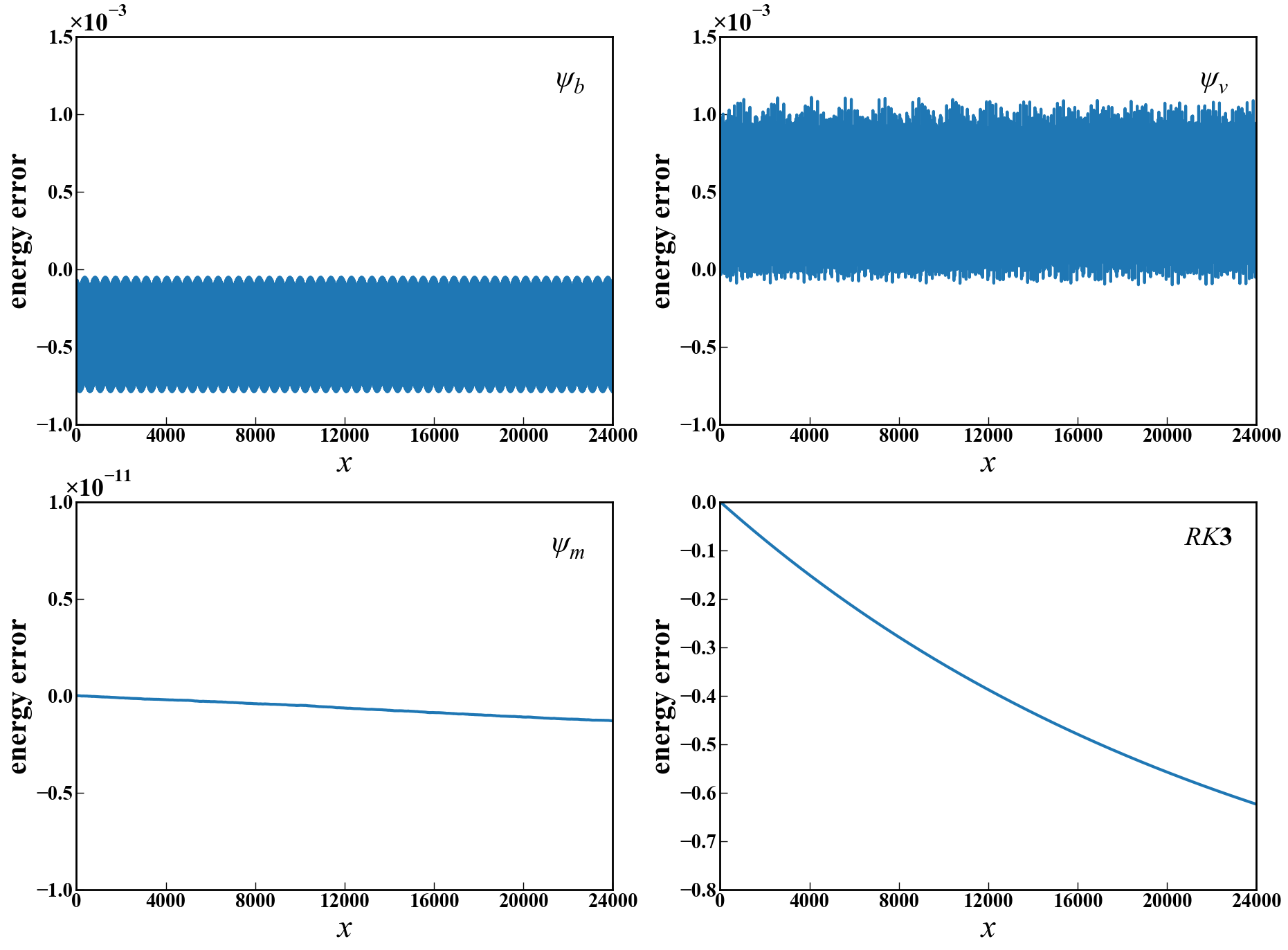}
  \caption{Relative errors of the energy $(E(n\Delta t)-E_0)/E_0$ are given by $\psi_b, \psi_v$, the midpoint scheme $\psi_m$, and RK3 over the time interval $[0,24000]$.}
  \label{fig1.2}
\end{figure*}

In panel (a) of Fig.~\ref{fig1.1} , we find that the numerical integrators $\psi_b, \psi_v$, the midpoint scheme $\psi_m$, and RK3 all give the accurate orbit in the first rotation period. In panel (b) of Fig.~\ref{fig1.1}, $\psi_b, \psi_v$, the midpoint scheme $\psi_m$ provide the correct orbit in the 300th rotation period while RK3 fails. Relative errors of the energy  $(E(n\Delta t)-E_0)/E_0$ of the numerical methods $\psi_b, \psi_v$ and the implicit midpoint scheme $\psi_m$ are bounded, which is shown in Fig.~\ref{fig1.2}. In particular, the energy $E$ is a quadratic invariant along the flow of phase space. The implicit midpoint scheme $\psi_m$ preserve the energy $E$ exact. So the relative energy error $(E(n\Delta t)-E_0)/E_0$ of the implicit midpoint scheme $\psi_m$ is tiny.
These numerical results verified the properties of long-term  near-conservation of energy for $\psi_b$, $\psi_v$, and the midpoint scheme $\psi_m$ and demonstrate the property of symplecticity of $\psi_b, \psi_m$. Since RK3 have not such good properties, it is not surprising on the failure of RK3 in orbit scale and energy conservation.


\subsection{Earth-Moon system}
The motion in the corotating coordinate is critical in the field of astronomy and space science. In the design of spacecraft orbit, the earth and moon disturbances should be taken into account when calculating the orbits of satellites near the earth and moon. In this case, the motions of the earth, moon and satellite form a restricted three body problem 
, which is a kind of corotating coordinate system.

The restricted three-body problem can be written in the form of (\ref{rotate}) with potential 
\begin{equation}
\begin{split}
U(\mathbf{x})=&-\dfrac{GM_1}{\sqrt{(x-x_1)^2+(y-y_1)^2+z^2}}\\
&-\dfrac{GM_2}{\sqrt{(x-x_2)^2+(y-y_2)^2+z^2}},
\end{split}
\end{equation}
which has been widely studied 
\citep{Gao2014AJ,Perdomo2017,ABOUELMAGD2020}. We expand into the component form of $(x,y,z)$, as follow 
\begin{equation}\notag
\begin{split}\label{eq:3.4.1}
\frac{d^2x}{dt^2}\!-\!x\omega^2\!-\!2\omega\frac{dy}{dt}\!&= -\frac{GM_1(x-x_1)}{R_1^{3}}\!-\!\frac{GM_2(x-x_2)}{R_2^{3}},\\
\frac{d^2y}{dt^2}\!-\!y\omega^2\!+\!2\omega\frac{dx}{dt}\!&= -\frac{GM_1(y-y_1)}{R_1^{3}}\!-\!\frac{GM_2(y-y_2)}{R_2^{3}},\\
\frac{d^2z}{dt^2}\!&= -\frac{GM_1z}{R_1^{3}}\!-\!\frac{GM_2z}{R_2^{3}},
\end{split}
\end{equation}
where $R_1=((x-x_1)^2+(y-y_1)^2+z^2)^{1/2}, R_2=((x-x_2)^2+(y-y_2)^2+z^2)^{1/2}$ and the coordinate origin is mass center of the system.

We study the Earth-Moon system. The unit of distance is an Astronomical Unit (1.4959787e13 cm), time unit is an earth day (86400 second) and mass unit is kilogram. The corresponding normalized parameters $GM_1=$ 0.8997011603631609e-09  (following the parameters in  \citep{Tu2020}), $GM_2=0.0123GM_1$, the distance of between earth and moon $R=$ 2.56267e-3, $r_1=-{M_2R}/{(M_1+M_2)}, r_2={M_1R}/{(M_1+M_2)}$ and the rotation speed $\omega = \sqrt{{G(M_1+M_2)}/{R^3}}$. The earth initial position is $(x_1,y_1)=(r_1,0)$ and the moon initial position $(x_2,y_2)=(r_2,0)$. We set two groups of initial conditions to check the behaviors of the above numerical integrators. 

{\bf Orbit 1}: we set initial position of a massless object at $\mathbf{x}=(-r_2/4,0,0)$, velocity $\dot{\mathbf{x}}=$ (0,1.69561e-3,0) and step-size $\Delta t = 0.01$. The orbit of the celestial object is numerically integrated over the time interval of $[0,40000]$. The energy of the system is a conserved quantity, $E_0=-7.1941456034028234e-8$.

{\bf Orbit 2}: we set initial position $\mathbf{x}=(-{3}r_2/5,0,0)$, velocity $\dot{\mathbf{x}}=$ (0,1.35057e-3,0) and step-size $\Delta t = 0.04$. The orbit is  integrated during time $t\in[0,10^5]$. The energy of the system is $E_0=2.4213436261924679e-7$. 
\begin{figure*}[htb]
  \centering
  \includegraphics[width=0.45\linewidth]{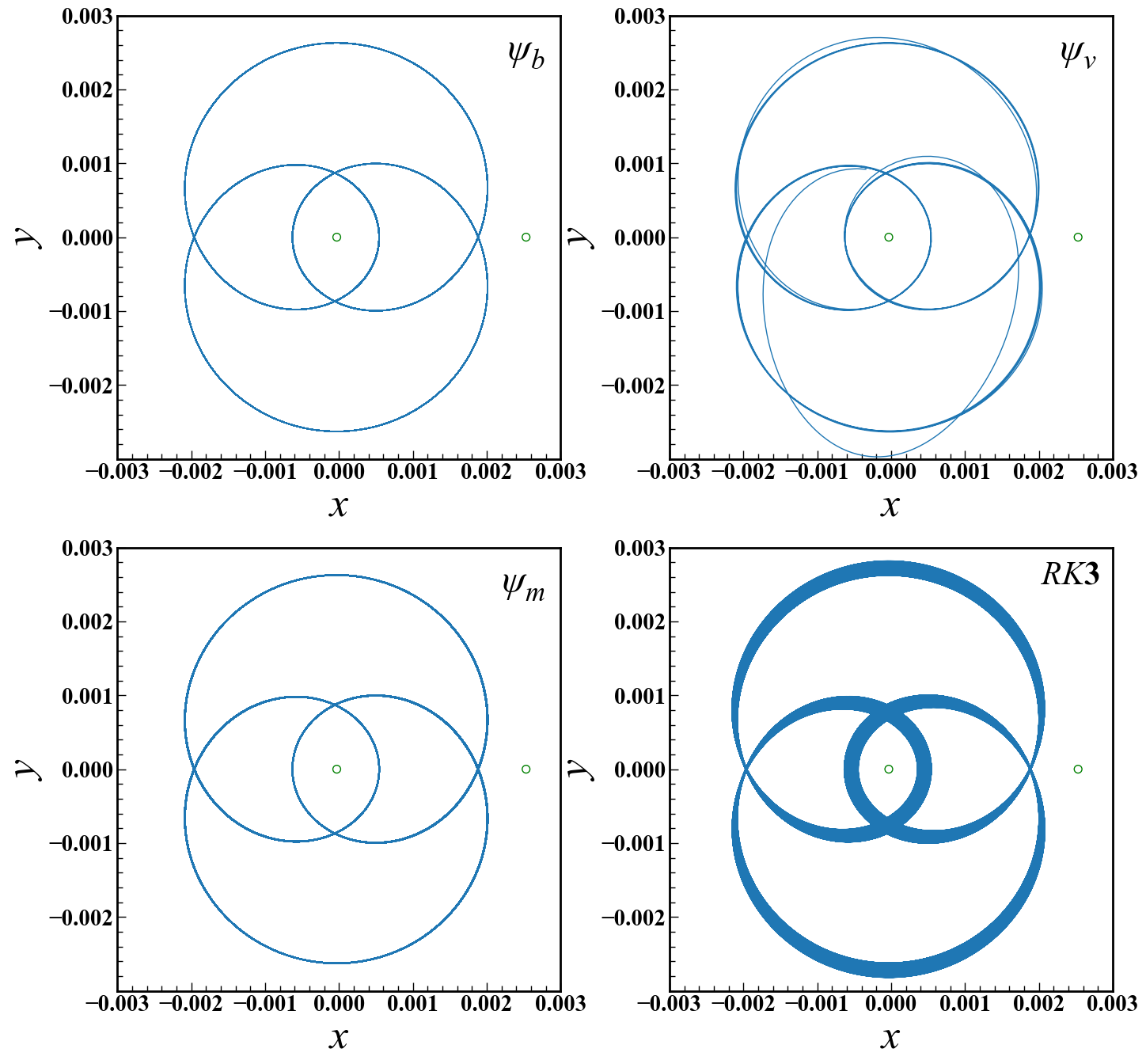} 
  \includegraphics[width=0.45\linewidth]{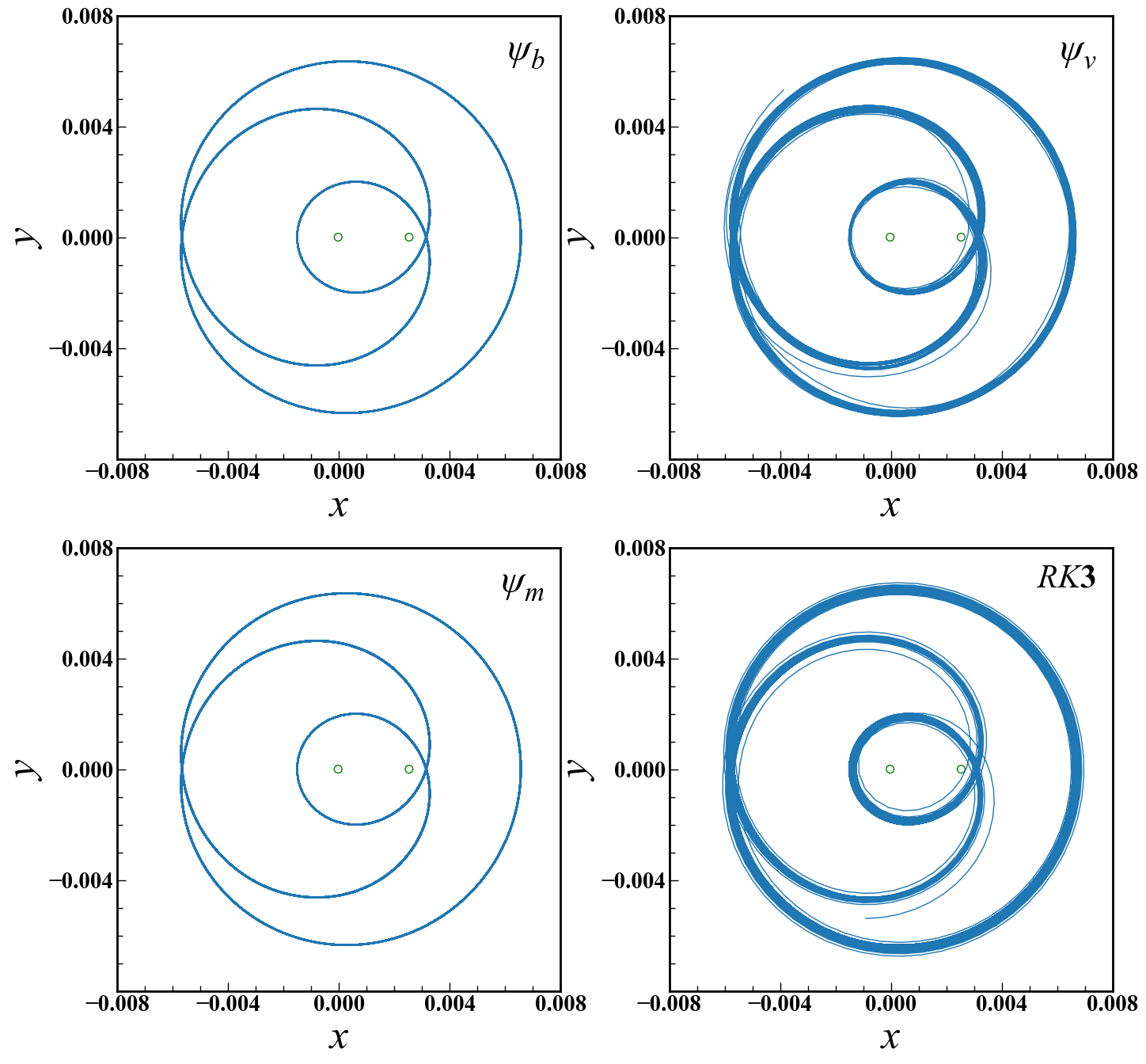} 
  \caption{Numerical orbits are given by $\psi_b, \psi_v$, $\psi_m$ and RK3 of orbit 1 (left) and orbit 2 (right), respectively.}
  \label{fig2.1}
\end{figure*}

\begin{figure*}[htb]
  \centering
  \includegraphics[width=0.8\linewidth]{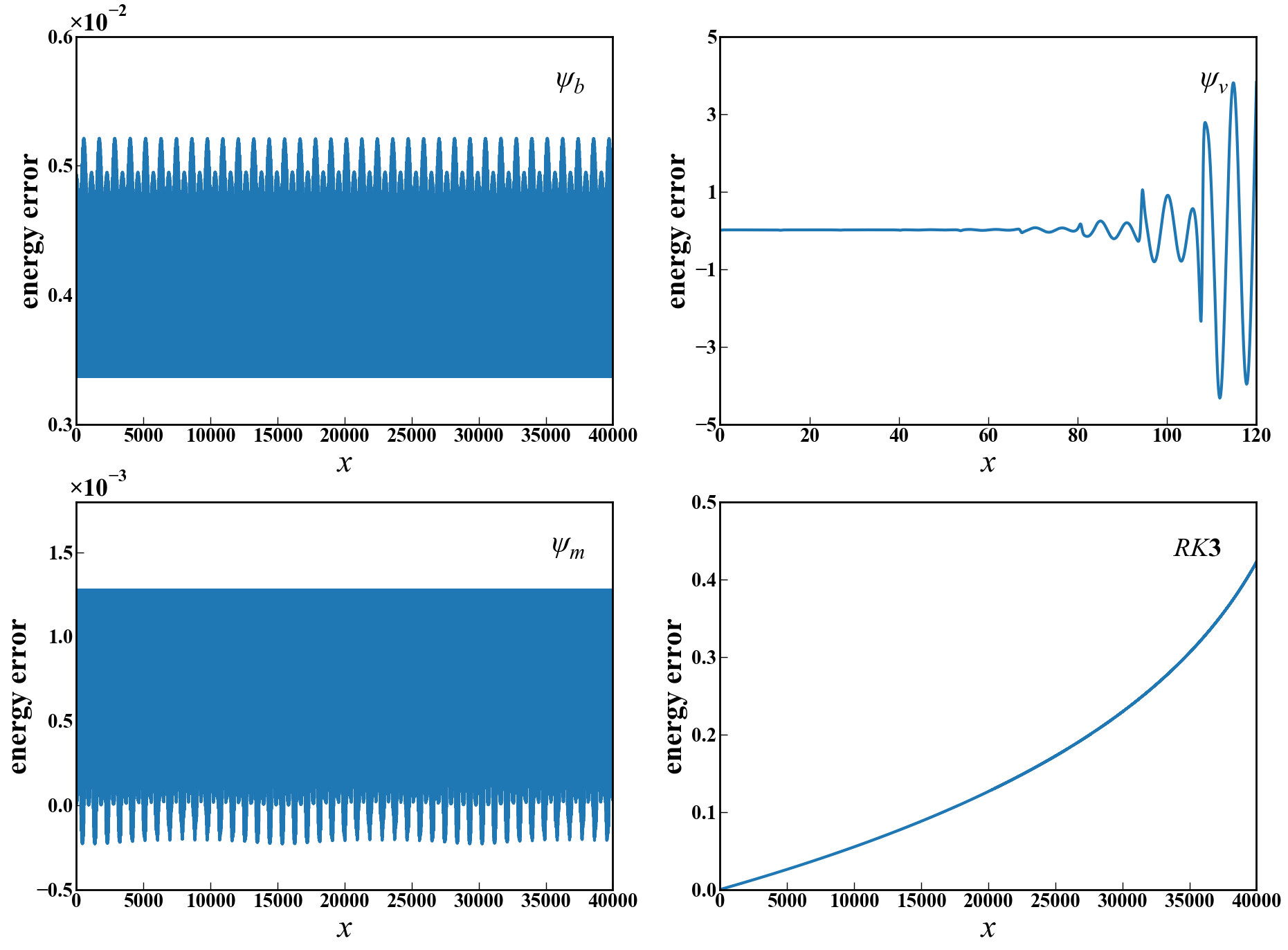}
  \caption{Relative errors of the energy $(E(n\Delta t)-E_0)/E_0$ are given by $\psi_b, \psi_v$, $\psi_m$, and RK3 of orbit 1 over the time interval $[0,40000]$.}
  \label{fig2.2}
\end{figure*}

\begin{figure*}[htb]
  \centering
  \includegraphics[width=0.8\linewidth]{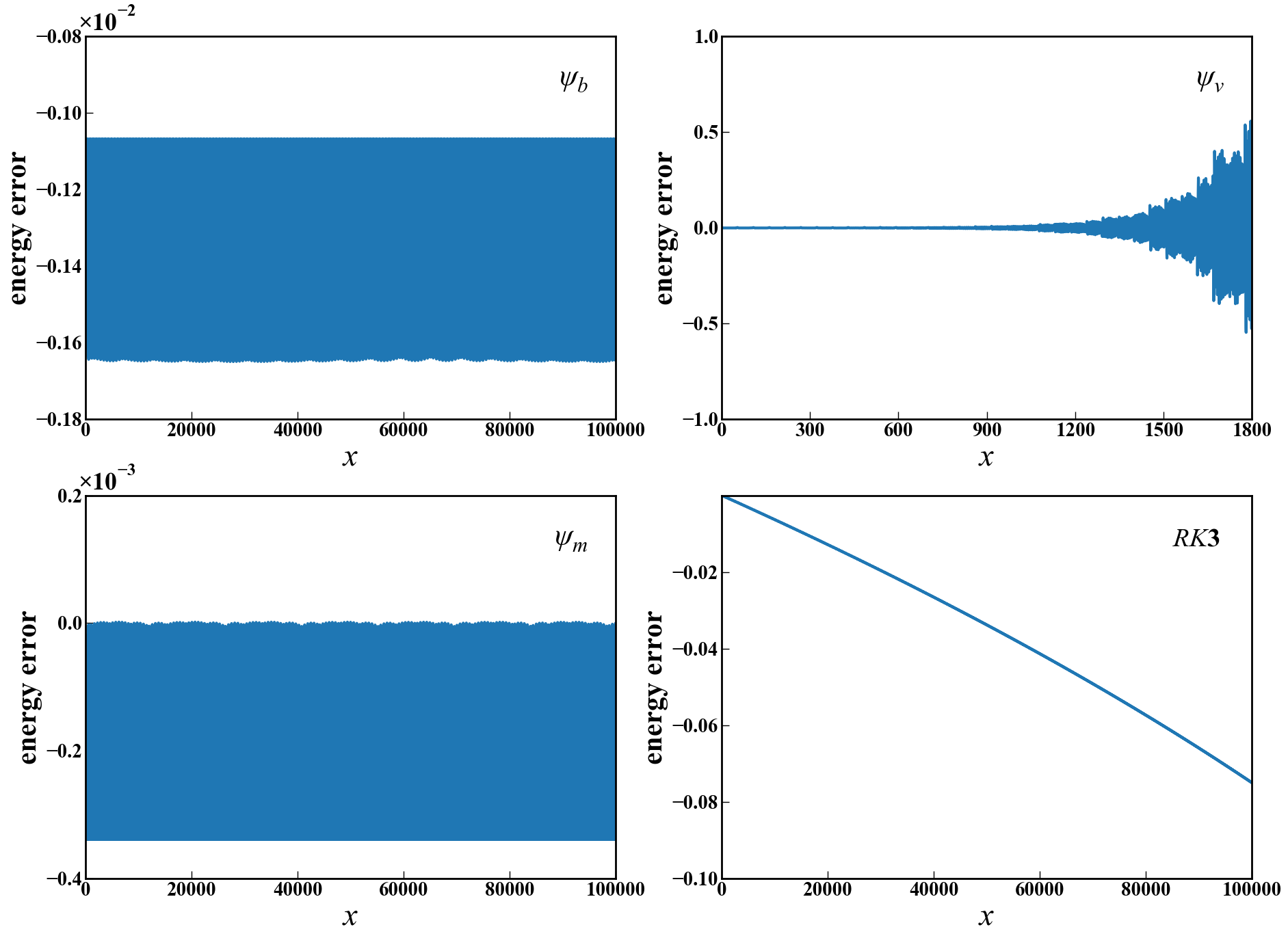}
  \caption{Relative errors of the energy $(E(n\Delta t)-E_0)/E_0$ are given by $\psi_b, \psi_v$, $\psi_m$, and RK3 of orbit 2 over the time interval $[0,10^5]$.}
  \label{fig2.3}
\end{figure*}

In Fig.~\ref{fig2.1}, we find that the numerical integrator $\psi_b$ and $\psi_m$ give the correct orbits over long time, but $\psi_v$ and RK3 fail. Fig.~\ref{fig2.2} and Fig.~\ref{fig2.3} present that the relative energy error $(E(n\Delta t)-E_0)/E_0$ with respect to the initial values of two orbits, respectively. In both case, $\psi_b$ and $ \psi_m$ perserve near-conservation of energy over long time, due to symplectic. In contrast, the error of $\psi_v$ and RK3 diverge. 
Note that non-symplectic $\psi_v$ actually maintains the property of near-conservation of energy, but it still behaves bad.

\begin{figure*}[htb]
  \centering
  \includegraphics[width=0.48\linewidth]{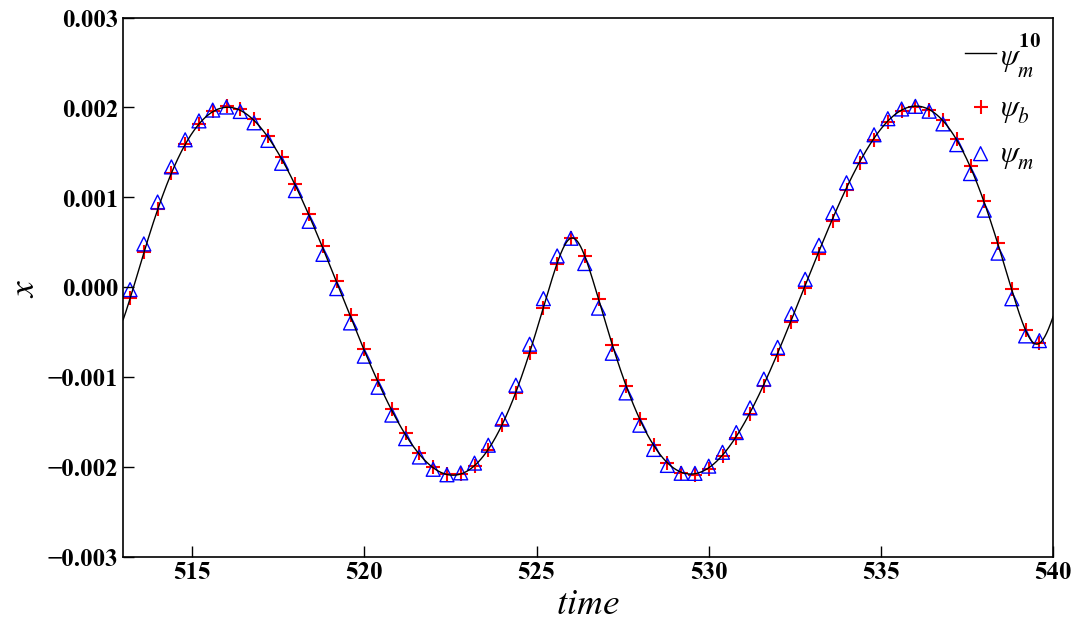}
  \includegraphics[width=0.48\linewidth]{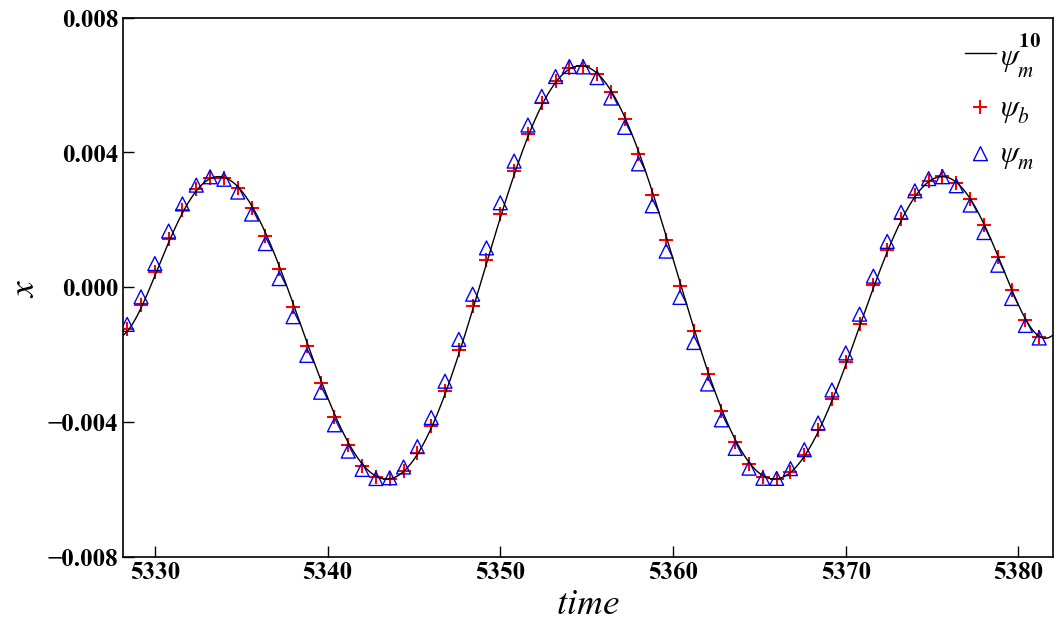}
  \caption{Numerical solutions in the x-direction of orbit 1 in the 20th orbit period (the left panel) and orbit 2  in the 100th orbit period  (the right panel).}
  \label{fig:6}
\end{figure*}

Besides the conservation of energy, we check the phase-drifting of the orbits for $\psi_b$ and $\psi_m$. Fig.~\ref{fig:6} shows numerical solutions of the 20-th and 100-th orbit period in the x-direction with respect to the initial values of two orbits, respectively. The $\psi_m^{10}$ denotes a 10th order composition method (formula (17) in \citet{Sofroniou2005}) of midpoint scheme and we uses 10 times finer step-size than the $\psi_b$ and $\psi_m$.

Comparing with the integrator $\psi_m$, we find that the phase of $\psi_b$ and $\psi_m$ are consistent with the high-order method $\psi_m^{10}$ and $\psi_b$ is slightly better than $\psi_m$. A reasonable speculation is that the integrator $\psi_b$ is explicit and there is less rounding error accumulation from the iterations than an implicit $\psi_m$.


\section{Conclusion}
\label{sect:conclusion}
In this paper, we investigated the symplectic property of three integrators, $\psi_b$, $\psi_v$ and $\psi_m$ in corotating coordinates. All of them are near-conservation of energy for long-term evolution and $\psi_b$ and $\psi_m$ are proved as sympletic schemes. In particular, the integrators of $\psi_b$ and $\psi_m$ are variational sympletic by directly discreting the motion equation and non-canonical Hamiltonian system, respectively.

Two groups of numerical experiments, rotating quadratic potential and earth-moon system, are carried out to verify our theoretical analysis. The energy error of $\psi_b$ and $\psi_m$  is indeed bounded and the phase shift also behaves well. However, the scheme $\psi_v$ is theoretically a conservative scheme, but it fails in phase-space evolution.

\normalem
\begin{acknowledgements}
We acknowledge the support from National SKA Program of China (Grant No. 2020SKA0110401), National Natural Science Foundation of China (Grant No. 11988101, 12171466), Special Research Assistant Program of the Chinese Academy of Sciences and K.C.Wong Education Foundation.
\end{acknowledgements}

\bibliographystyle{raa}
\bibliography{ref}



\clearpage
\end{document}